\documentclass[sigconf]{acmart}

\usepackage{booktabs} 

\usepackage{balance}

\begin{document}
\title[Adopting Automated Bug Assignment]{Adopting Automated Bug Assignment in Practice\\- A Registered Report of an Industrial Case Study}

\author{Markus Borg}
\orcid{XXX}
\affiliation{%
  \institution{RISE Research Institutes of Sweden\\Lund University}
  \city{Lund}
  \country{Sweden}
}
\email{markus.borg@ri.se}

\author{Leif Jonsson}
\orcid{XXX}
\affiliation{%
  \institution{Ericsson}
  \city{Kista}
  \country{Sweden}
}
\email{leif.jonsson@ericsson.com}

\author{Emelie Engström}
\orcid{XXX}
\affiliation{%
  \institution{Lund University}
  \city{Lund}
  \country{Sweden}
}
\email{emelie.engstrom@cs.lth.se}

\author{Béla Bartalos}
\orcid{XXX}
\affiliation{%
  \institution{Ericsson}
  \city{Budapest}
  \country{Hungary}
}
\email{bela.bartalos@ericsson.com}

\author{Attila Szabo}
\orcid{XXX}
\affiliation{%
  \institution{Ericsson}
  \city{Budapest}
  \country{Hungary}
}
\email{attila.szabo@ericsson.com}

\renewcommand{\shortauthors}{M. Borg et al.}

\begin{abstract}
[Background/Context] The continuous inflow of bug reports is a considerable challenge in large development projects. Inspired by contemporary work on mining software repositories, we designed a prototype bug assignment solution based on machine learning in 2011-2016. The prototype evolved into an internal Ericsson product, TRR, in 2017-2018. TRR’s first bug assignment without human intervention happened in 2019. [Objective/Aim] Our exploratory study will evaluate the adoption of TRR within its industrial context at Ericsson. We seek to understand 1) how TRR performs in the field, 2) what value TRR provides to Ericsson, and 3) how TRR has influenced the ways of working. Secondly, we will provide lessons learned related to productization of a research prototype within a company. [Method] We design an industrial case study combining interviews with TRR developers and users with analysis of data extracted from the bug tracking system at Ericsson. Furthermore, we will analyze sprint planning meetings recorded during the productization. Our data analysis will include thematic analysis, descriptive statistics, and Bayesian causal analysis.
\end{abstract}

\copyrightyear{2021} 
\acmYear{2021} 
\acmConference[ESEM '21]{Registered Report Submitted to the 15th ACM/IEEE International Symposium on Empirical Software Engineering and Measurement}{October 11--15, 2021}{Bari, Italy}

%
%



\maketitle

\section{Introduction} \label{sec:intro}
In large development projects, the continuous inflow of bug reports is a considerable challenge~\cite{bettenburg2008duplicate,just2008towards}. The Bug Tracking System (BTS) is a central repository in contemporary software development organizations. There are two archetypal bug assignment processes, i.e., approaches to distribute bug reports to developers. First, as is common in Open-Source Software (OSS) communities, individual developers can select bug reports to resolve in a \textit{pull-based process}. Second, a \textit{push-based process} can be used where a change control board or product manager assigns bug reports to either development teams or individual developers. In our research, we focus on the latter, i.e., push-based bug assignment to development teams.

Push-based bug assignment is normally done manually. However, several studies report that manual bug assignment is labor-intensive and error-prone~\cite{baysal_bug_2009,jeong2009improving}, resulting in ``bug tossing''~\cite{anvik2011reducing,bhattacharya2012automated,jonsson2012towards} and potentially slower bug resolution. Several researchers have proposed mitigating the challenges by automating bug assignment. The most common automation approach uses supervised Machine Learning (ML), i.e., a classifier is trained to find patterns in historical bug reports to make recommendations for new bugs. Early research on automated bug assignment focused on OSS development communities, especially the Eclipse and Mozilla projects. However, the OSS context differs from proprietary development in several aspects, e.g., organizational structures and developer incentives. A recent study at LGE Brazil constitutes a rare example of an empirical study in a large company~\cite{oliveira2021issue}. 

In 2016, we presented a controlled experiment on ML-based bug assignment using five datasets from two companies in telecommunications and process automation~\cite{jonsson2016automated}. This study was the first step in an incremental design science research process~\citep{engstrom2020software}. Our findings in this controlled setting were positive and led to internal productization of a simplified version of the solution within Ericsson. Since 2017, a team in Hungary owns and maintains the solution, referred to as Trouble Report Routing (TRR). To align the terminology, we refer to bug reports as Trouble Reports (TR) in the remainder of this report.

We have previously reported lessons learned from deploying TRR in an anecdotal manner~\cite{carver2018industry}. Furthermore, we conducted a quantitative analysis of the prediction accuracy of TRR's assignments~\cite{sarkar2019improving}. In the latter paper, we concluded that the results were promising, but not yet accurate enough for everyday use in its intended target environment. We continued improving and customizing TRR and in 2019 activated the solution -- the very first TR assignment without human intervention happened on April 10, 2019. Since then, TRR has been in continuous operation and automatically routed roughly 30\% of the incoming TRs.

We are now designing an industrial case study to evaluate the adoption of TRR within its industrial context. Our study will present new perspectives on automated bug assignment in proprietary contexts by moving beyond the prediction accuracy that has been in focus in previous work~\cite{sarkar2019improving,oliveira2021issue}. Our study aims to provide insights regarding direct as well as indirect effects of deploying this research-based intervention in an operational setting. Thus, we will add empirical support for (as well as refinements of) the previously proposed \emph{technological rule}~\cite{runeson2020design}:
\begin{quote}{To achieve more effective assignment of bugs to teams in large scale industrial contexts, use ensemble-based machine learning to automate bug assignment.}\end{quote}
In the initial proposal, specifically ensemble-based ML approaches were recommended. However, the deployed version needed adaptation to the context and thus our starting point in this study is the more general technological rule:
\begin{quote}{To achieve more efficient and effective assignment of bug reports to teams in large scale industrial contexts, use machine learning to automate bug assignment.}\end{quote}

This registered report constitutes our case study protocol, developed in line with guidelines by Runeson \textit{et al.}~\cite{runeson2012case}. A summary of the elements of the research design is presented below, which also presents the structure of this report.

\begin{itemize}
    \item \textbf{Rationale} Evaluate the adoption of TRR within its industrial context. (Section~\ref{sec:ratpurp})
    \item \textbf{Purpose} Provide evidence on the industrial applicability of ML-based bug assignment. (Section~\ref{sec:ratpurp})
    \item \textbf{The case} Automated bug assignment using TRR at Ericsson. (Section~\ref{sec:caseunits})
    \item \textbf{Units of analysis} The team maintaining TRR, engineers conducting TR assignments, and two teams using TRR with different levels of automation. (Section~\ref{sec:caseunits})
    \item \textbf{Theory} The design science paradigm~\cite{runeson2020design} constitutes the frame for our research where the general technological rule is a proposition for efficient and effective bug assignment~\cite{jonsson2016automated}. In this study we add design knowledge related to the industrial adoption of this proposition. Our evaluation is guided by models related to quality~\cite{regnell_supporting_2008}, automation~\cite{parasuraman2000model}, and technology acceptance~\cite{davis1989perceived}. (Section~\ref{sec:theory})
    \item \textbf{Research Questions} Four RQs targeting different aspects of adoption of TRR:  1) evolution from prototype, 2) prediction accuracy, 3) added value, and 4) direct and indirect effects. (Section~\ref{sec:rqs})
    \item \textbf{Data collection} Quantitative data from the BTS and TRR. Qualitative data from interviews, recorded sprint meetings, and internal documentation. (Section~\ref{sec:collection})
    \item \textbf{Data analysis} Descriptive statistics complemented by causal Bayesian analysis~\cite{hernan2020causal}. Thematic analysis~\cite{cruzes_recommended_2011} to interpret the qualitative data. (Section~\ref{sec:analysis})
    \item \textbf{Quality assurance} Prolonged industry-academia collaboration to ensure relevance~\cite{garousi2020practical}. Rigor assured by method and researcher triangulation with member checking. (Section~\ref{sec:quality})
\end{itemize}

\section{Research Design and Goals}
\begin{figure}
    \centering
    \includegraphics[width=0.35\textwidth]{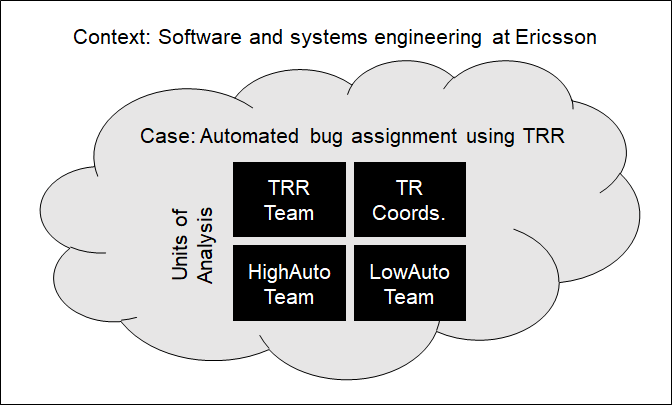}
    \caption{The context, case, and units of analysis.}
    \label{fig:contextcaseunits}
\end{figure}

We conduct interpretivist research as the methods of natural science are insufficient for understanding the case in its social reality context~\cite{baltes2020sampling}. Figure~\ref{fig:contextcaseunits} illustrates the context, the case under study, and the units of analysis. As defined by Runeson \textit{et al.}~\cite{runeson2012case}, \textit{``case study in software engineering is an empirical enquiry that draws on multiple sources of evidence to investigate one instance (or a small number of instances) of a contemporary software engineering phenomenon within its real-life context, especially when the boundary between phenomenon and context cannot be clearly specified.''} 

Since the adoption of TRR cannot be isolated from the development context at Ericsson, we design an industrial case study. Our study relies on a flexible design, i.e., the sampling, data collection as well as the data analysis involve components relying on our evolving knowledge about the phenomenon. This registered report presents how we design for flexibility while maintaining rigor.

\subsection{Rationale and Purpose} \label{sec:ratpurp}
Our overall goal is to evaluate the adoption of TRR within its industrial context at Ericsson (cf. Figure~\ref{fig:rqs}). Several aspects motivate us to pursue this goal. First, we want to follow up on research that was initiated 10 years ago. How does the automated bug assignment solution actually perform in the field? Are the assignments provided by TRR sufficiently accurate to provide value in the industrial context? Do engineers at Ericsson appreciate the support provided by TRR? How has the introduction of TRR influenced the ways of working? Are there any surprising indirect effects that should be reported? As discussed in Section~\ref{sec:intro}, there is a lack of industrial case studies sharing these types of insights. 

Second, we seek to provide insights regarding the industrial adoption of a research prototype. By conducting this study, we will highlight an example of industry-academia collaboration and technology transfer. The study will contain a retrospective analysis of the evolution from prototype to internal product. We will explore obstacles experienced in the productization and share lessons learned on how they were tackled in the industrial context. We expect our findings to be highly relevant for other software engineering researchers proposing new tools for use in proprietary contexts.

\subsection{Context} \label{sec:context}
As illustrated in Figure~\ref{fig:contextcaseunits}, the context is software and systems engineering at Ericsson. Ericsson is a global actor in telecommunications. We characterize the context inspired by the facets proposed by Petersen \textit{et al.}~\cite{petersen_context_2009}, focusing on the factors that we believe are the most relevant for our study.

\textbf{Product} The products in the analysis consists of two large systems in the Information and Communications Technology (ICT) domain. Various programming languages are used in the products, but a majority of the code is developed in C++ and Java. Other languages such as hardware description languages and tailored domain-specific languages are also used. The two systems are mature with old code bases.

\textbf{Processes} The project model used to develop both systems is an adapted agile development process. Development in the ICT domain is heavily standardized, and adheres to standards by regulatory bodies such as 3GPP, 3GPP2, ETSI, IEEE, IETF, ITU, and OMA. Moreover, Ericsson is ISO~9001 and TL~9000 certified. 

\textbf{Practices and Techniques} The development projects use agile practices that have been customized for the organization, e.g., sprint planning meetings, retrospectives, self-organization, and test automation. The development projects are organized into two-week sprints followed by releases. 

\textbf{People} Staff turnover is very low in the development organization. Many of the engineers are seniors developers who have been working on the same, or similar, products for many years.

\textbf{Organization} Several hundreds of engineers distributed over several countries, e.g., Sweden, Hungary, China, and Canada. In total, Ericsson has 100,000 global employees. The BTS is the central point for organizing the bug handling process. Tracking of analysis, implementation proposals, testing, and verification are all coordinated through the BTS.

\textbf{Market} Both systems are deployed at customer sites world-wide in the ICT market. The telecommunications market is currently in a transition from the last generation of 4G networks to 5G. Software-oriented technology improvements are increasingly flexible high-speed connectivity at ultra-low latency.

\subsection{Case and Units of Analysis} \label{sec:caseunits}
\begin{figure}
    \centering
    \includegraphics[width=0.40\textwidth]{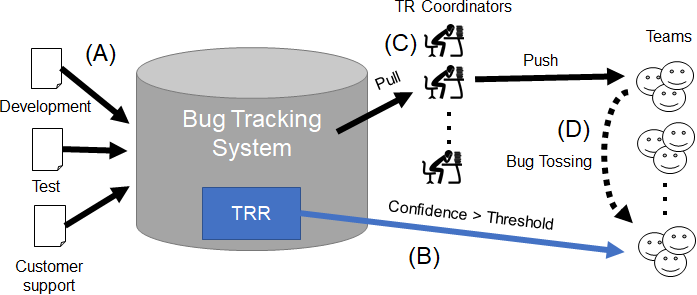}
    \caption{TRR in operation in the industrial context.}
    \label{fig:mlaflcontext}
\end{figure}

The case under study is automated bug assignment using TRR in its industrial context. Figure~\ref{fig:mlaflcontext} shows how TRR has been integrated in the BTS at the company. Different organizational units submit TRs to the BTS (A). TRR, operating as a BTS plug-in, predicts which development team would be the most likely to resolve the bug and appends this information to the TR. If the prediction has a high confidence value, i.e., above a configurable threshold, the TR is automatically assigned to the corresponding team (B). If the confidence value is lower than the threshold, the assignment process relies on the normal manual approach by one of the TR coordinators (C). The manual approach encompasses a TR coordinator pulling a TR from the BTS, analyzing it (possibly guided by TRR's recommendation~\cite{borg2014changes}), and pushing the TR to one of the development teams. Bug tossing entails reassignment of a TR to another team (D). Note that the phenomenon of bug tossing is not necessarily caused by an incorrect initial team assignment. On the contrary, it can be a required step when resolving complex bugs that necessitate changes by multiple teams.

TRR is currently in operation in a BTS used for development of two major systems consisting of 19 subsystems. The subsystems are developed by corresponding virtual organizational units -- in this study we simply refer to them as ``teams'' for brevity. Nine of the 19 teams have opted in to receive TRs automatically assigned by TRR. For the remaining 10 teams, TRR only attaches its prediction to the TRs as a recommendation. According to Parasuraman \textit{et al.}'s model of automation~\cite{parasuraman2000model}, the automatic vs. recommended TR assignment correspond to automation level 8 and level 4, respectively. We plan to use the different levels of TRR automation for comparisons and refer to them as ``HighAuto'' and ``LowAuto''. 

TRR is since 2017 maintained by a team in Hungary, see ``TRR Team'' listed as a unit of analysis in Figure~\ref{fig:contextcaseunits}. Furthermore, we define three additional units of analysis. First, a development team that opted in as early adopters of the HighAuto TRR, heavily involved in the transition from research prototype to operational tool (cf. ``HighAuto Team''). Second, a development team that opted out from automatic routing, i.e., representing LowAuto TRR. Third, engineers that act as TR coordinators for either teams using HighAuto TRR, LowAuto TRR or not using TRR at all (cf. ``TR Coords.''). TR coordinators have different roles within Ericsson, but perform TR assignment as part of their routine work.

\subsection{Theory} \label{sec:theory}
The general \emph{problem} of inefficient and ineffective bug assignment was observed in the literature~\cite{bettenburg2008duplicate,just2008towards,oliveira2021issue} as well as in the specific industrial contexts where this research was conducted~\cite{jonsson2012towards,sarkar2019improving}. With the solution in mind (to use ML techniques to assign TRs to teams), the characteristics of the targeted \emph{problem instance} were identified, i.e., we explored the nature of the TRs, the BTS, and the organizational context within a subset of the development at Ericsson. Related work on bug classification as well as on ML techniques was identified~\citep{jonsson2016automated}, which underpinned the \emph{design decisions} for the proposed solution. The ML solutions were implemented and trained using the Weka framework~\citep{hall_weka_2009}. Several alternative solution instances were \emph{validated} on real data (50,000 TRs) from five projects across two companies/domains. For the specific companies, a design artifact was produced, namely a prototype ensemble-based bug assignment tool built on top of Weka.

\begin{figure}
    \centering
    \includegraphics[width=0.21\textwidth]{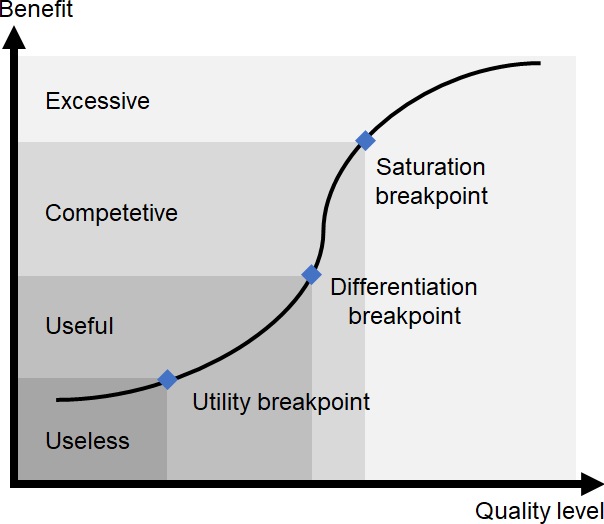}
    \caption{The benefit/quality relation in the QUPER model.}
    \label{fig:quper}
\end{figure}

In our 2016 paper, we stated that the translation from TRR's prediction accuracy to the practical value of the solution might not be linear. Furthermore, we discussed this aspect in terms of the QUPER model~\cite{regnell_supporting_2008}, a theoretical construct describing the perceived benefits of different degrees of quality as continuous and non-linear. Figure~\ref{fig:quper} shows the three quality breakpoints proposed by the QUPER model for TRR:
\begin{itemize}
    \item \textbf{Utility} Engineers start considering TRR as a useful addition to manual bug assignment.
    \item \textbf{Differentiation} Engineers recognize that TRR provides a competitive advantage compared to fully manual work.
    \item \textbf{Saturation} Increasing the quality of TRR beyond this points adds no practical value.
\end{itemize}

We will base our evaluation of the adoption of TRR on three theoretical models. First, we will revisit the QUPER model to assess where TRR belongs on the sliding quality scale. Second, as we also did in the original paper, we will discuss the increased level of automation using the model by Parasuraman \textit{et al.}~\cite{parasuraman2000model}. The latter model opens up for an analysis of both direct and indirect effects of increased automation. Third, we will study the Ericsson engineers' impressions of working with TRR from the perspective of the established Technology Acceptance Model (TAM)~\cite{davis1989perceived}. 

\subsection{Research Questions} \label{sec:rqs}
\begin{figure}
    \centering
    \includegraphics[width=0.5\textwidth]{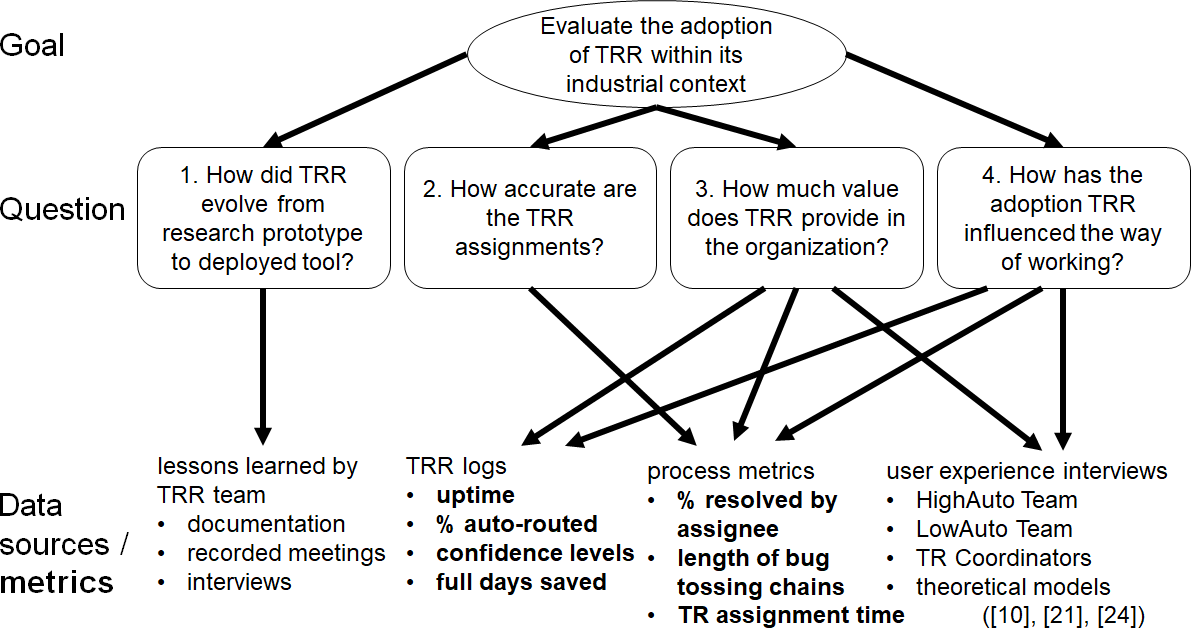}
    \caption{Breakdown of our goal to research questions, data sources, and metrics.}
    \label{fig:rqs}
\end{figure}

As visualized in Figure~\ref{fig:rqs}, the aim of the study is to evaluate the adoption of TRR within its industrial context. We have defined four main research questions, which may all be answered by applying both qualitative and quantitative methods. The lower part of the figure presents data sources and metrics, where the latter are indicated in bold font.
\begin{itemize}
\item[RQ1] How did TRR evolve from prototype to deployed tool?
\item[RQ2] How accurate are the TRR assignments?
\item[RQ3] How much value does TRR provide in the organization?
\item[RQ4] How has the adoption TRR influenced the way of working?
\end{itemize}

RQ1 will be answered by studying the design decisions Ericsson engineers made along the way. How and why were potential adaptations to the original solution made? What were the major challenges during the tool introduction, including processes, technology, organizational issues, and human factors? The TRR team, one of four units of analysis, will share a collection of recorded virtual sprint meetings and internal documentation. Furthermore, we will conduct interviews to collect lessons learned.

RQ2 involves a quantitative analysis of TRR's prediction accuracy in the light of our previous work~\cite{sarkar2019improving}. Previously, we studied the accuracy relying on a set of roughly 10,000 TRs. Relying on easily accessible textual and categorical features, we obtained precision and recall values around 80~\%. As this was reported as insufficiently accurate for regular use, we proposed to only assign TRs for which the ML classifier was confident. We will now revisit the accuracy RQ to evaluate how TRR performed in the field using historical data since deployment in April 2019, incl. the fraction of TRs resolved by the first assigned team and the length of bug tossing chains.  

RQ3 targets the utility of TRR and its added value in the organization. We will complement the insights provided by RQ2 with an analysis of the TRR utilization, i.e., whether it has been available (uptime) and sufficiently confident to be effective (fraction of automatic TR assignments and distribution of confidence levels). Moreover, we will complement the analysis with qualitative insights from interviews with members of the HighAuto and LowAuto Teams and a sample of TR coordinators (cf. Figure~\ref{fig:contextcaseunits}). Section~\ref{lab:collQual} presents how we will design interviews supported by theoretical models~\cite{regnell_supporting_2008,parasuraman2000model,davis1989perceived}. Analyzing differences between HighAuto TRR and LowAuto TRR will enable comparisons.

RQ4 explores the direct and indirect effects of introducing TRR in the organization. A tool never exists in isolation, i.e., the introduction of tool-oriented interventions ought to be studied through a holistic perspective. Among other things, we seek to understand what made certain teams opt-in to the HighAuto TRR whereas others preferred LowAuto TRR. Analogous to RQ3, RQ4 will be answered using a combination of quantitative metrics and rich information from interviews.

\section{Execution Plan}
\begin{figure}
    \centering
    \includegraphics[width=0.3\textwidth]{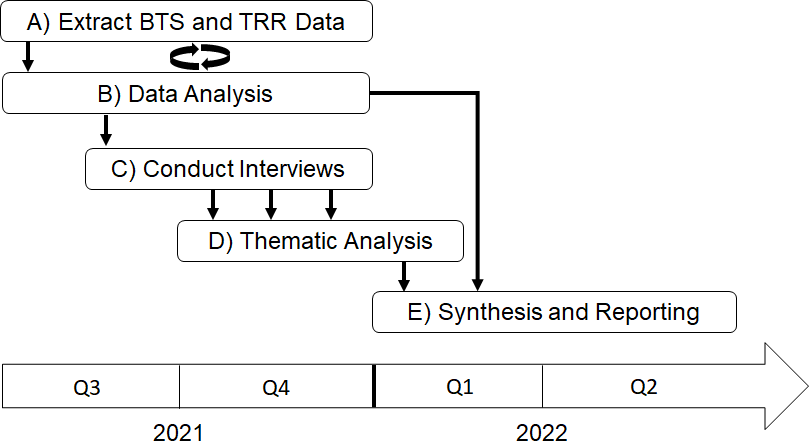}
    \caption{Primary activities during the study execution.}
    \label{fig:plan}
\end{figure}

Figure~\ref{fig:plan} shows an overview of the execution plan. Section~\ref{sec:collection} describes how we will proceed with data collection during Q3-Q4 2021. Subsequently, Section~\ref{sec:analysis} presents our approach to data analysis. Interviews will be conducted in Q3-Q4 2021 and the corresponding analysis will be concluded during Q1 2022. Quantitative analysis will be done during Q3-Q4 2021. Research synthesis will be initiated in Q1 2022 and the reporting will be concluded in Q2 2022.

\subsection{Data Collection} \label{sec:collection}
The study relies on non-probability sampling~\cite{baltes2020sampling}, i.e., there will be no element of randomness when selecting items in the sampling frame. Instead, we will use a combination of purposive and referral-chain sampling to select interviewees. To mitigate selection bias, our initial set of interviewees will include engineers from different levels of the organisation as well as with varying levels of adoption of TRR. Furthermore, we will add questions to the interview guide with the purpose of identifying people having complementary insights. For our artifact analyses, i.e., document analysis and mining software repositories, we will use whole-frame sampling.

\subsubsection{Quantitative Data}
The BTS is an important source of data that constitutes a valuable target for mining of software repositories~\cite{borg2014changes}. The BTS data contain details of TRs, e.g., assignments, submitters, severity levels, and time stamps. We plan to collect all data related to the development of the 19 teams that either use HighAuto TRR or LowAuto TRR, (cf. A) in Figure~\ref{fig:plan}). We will collect 2,5 years' worth of data, i.e., 2019-04-10--2021-10-10. Thus, we apply \textit{whole-frame sampling} by selecting all items in the sampling frame~\cite{baltes2020sampling}. TRR logs all its actions and output in the BTS. The logs primarily show the TRR predictions, i.e., the bug assignment output provided by the tool. Furthermore, the logs contain confidence levels accompanying the predictions. Finally, all TRR actions have individual time stamps.

\subsubsection{Qualitative Data} \label{lab:collQual}
We will select interviewees from the four units of analysis based on \textit{purposive sampling}. Our goal is to identify the candidate interviewees that can provide the richest information, while also complementing the perspectives of the previous interviewees from a heterogeneity perspective, e.g., roles, background, site, age, and gender. As case study research allows a flexible design, we will complement the interviewee selection with \textit{referral-chain sampling}. In practice, each interview session will conclude by asking the interviewee to refer other members of the population whom they believe would provide valuable perspectives on the adoption of TRR. End of sampling will be determined by the level of saturation reached with respect to the initial coding in the thematic analysis.

We will develop an interview guide with some variation points for the four units of analysis. The interview sessions with TRR Team, HighAuto Team, and LowAuto Team will focus on challenges,  solutions, and opportunities related to the evolution of TRR from a research prototype to an internal tool at Ericsson. On the other hand, the interview sessions with the TR Coordinators will primarily focus on the user experience and perceived value of TRR (corresponding to perceived ease of use and usefulness in TAM~\cite{davis1989perceived}). However, we anticipate that the interview questions will be intermixed in several interview sessions, i.e., we will perform semi-structured interviews. An initial overview of the interview guide is presented below:

\begin{enumerate}
    \item A formal introduction including overall purpose, non-disclosure agreements, integrity, security, and research ethics.
    \item A brief description of the interviewee's current role and engineering background.
    \item (If applicable) Lessons learned related to evolving TRR into an internal tool.
    \item (If applicable) Perceived TRR value and ease of use (guided by TAM as done for testing tools by Mezhuyev \textit{et al.}~\cite{mezhuyev2019acceptance}).
    \item (If applicable) Perceived TRR value in relation to its prediction accuracy (supported by descriptive statistics and the QUPER model~\cite{regnell_supporting_2008}, in line with our previous work on change impact analysis~\cite{borg2016supporting}).
    \item Reflections on direct and indirect effects when increasing the level of automation in bug assignment (guided by Parasuraman \textit{et al.}~\cite{parasuraman2000model}).
    \item Final comments and suggestions for additional interviewees.
\end{enumerate}

Interview sessions are expected to last 30-60 min and will be conducted by at least two interviewers. All sessions will be done remotely using MS Teams as Ericsson engineers will be working from home during 2021 due to the Covid-19 pandemic. The three arrows from C) to D) in Figure~\ref{fig:plan} illustrates how data collected from the four units of analysis enter the thematic analysis.

\subsection{Data Analysis} \label{sec:analysis}
This subsection describes how we will analyze the BTS/TRR data and our approach to qualitative analysis.

\subsubsection{Quantitative Data} \label{sec:quan_anal}
We open the discussion on quantitative data analysis with an important disclaimer. Bug data is highly sensitive to any development organization. As a result, we will never be able to report any absolute numbers related to TRs. Instead, all bug counts will most likely be presented in relative numbers.

First, we will use the extracted data to calculate simple descriptive statistics for both HighAuto TRR and LowAuto TRR (cf. the leftmost arrow from A) in Figure~\ref{fig:plan}). The descriptive statistics will be used as input to the interview sessions (cf. the arrow from B) to C) in Figure~\ref{fig:plan}).

Second, we will iteratively extract data and conduct the corresponding analysis. The cycle between A) and B) in Figure\ref{fig:plan} highlights that this activity is partly exploratory, i.e., we expect to find new research angles as we get more familiar with the data. Our initial list of metrics, also presented in Figure~\ref{fig:rqs}, are presented below:

\begin{enumerate}
    \item \textbf{Uptime} will be estimated by calculating the fraction of TRs with missing TRR predictions. TRR is deployed as a BTS plug-in rather than a separate web service, thus we will estimate TRR service outage through missing output.
    \item \textbf{Fraction automatically routed} will be calculated from the TRR logs. Ericsson estimates that a manual TR assignment takes 2~min on average, i.e., we will naively report TR coordinators' potential time savings.
    \item \textbf{Distribution of confidence levels} for the TRR predictions will be collected from the TRR logs. The confidence level is fundamental as it must surpass a certain threshold to allow automated assignments.
    \item \textbf{Fraction of TRs resolved by the assigned team} will be calculated by combining BTS data and TRR logs. This represents an ideal case, i.e., the team assigned the TR also resolved it.
    \item \textbf{Average length of bug tossing chains} shows the number of TR reassignments. This measure is commonly reported in studies on automated bug assignment~\cite{jeong2009improving,wu2018empirical}.
    \item \textbf{Average time to assign TRs} will be calculated from the BTS data, i.e., the average time between TR submission and assignment.
    \item \textbf{Full days saved} is an in-house metric used by Ericsson for initial evaluations of TRR. TR Coordination meetings, i.e., manual TR assignment, are scheduled Mon-Fri in the morning hours CET. If a new TR is submitted shortly after this meeting, it would not be assigned until the meeting the next weekday -- TRR could then potentially save a full day.
\end{enumerate}

While we intend to compare the metrics for HighAuto TRR and LowAuto TRR, we will be conservative about making causal claims. There are two approaches to drawing causal conclusions, Randomized Controlled Trials (RCT) and Bayesian Causal Analysis (BCA)~\cite{hernan2020causal}. These are mathematically proven to be equivalent \cite{pearl2009causal}. Unfortunately, an RCT is not a viable option in the organization. To compensate, we will perform a BCA to try to detect a causal effect of opting in to HighAuto TRR vs. opting out (LowAuto TRR). However, we are aware that our qualitative analysis might reveal unmeasurable confounding factors that invalidate HighAuto and LowAuto comparisons in our case under study. We will quantify these in a graphical model and measure the sensitivity to model noise and model misclassification as part of the BCA workflow. Moreover, the comparative descriptive statistics will stimulate discussions during the interview sessions in relation to the qualitative analysis of RQ3, i.e., the value of TRR within Ericsson.

\subsubsection{Qualitative Data}
To answer RQ1, RQ3, and RQ4 we will analyse documentation, recorded sprint meetings, and interview transcriptions. 
We will iterate over the five steps of thematic analysis as described by Cruzes and Dybå~\cite{cruzes_recommended_2011}: 1) extract relevant data, 2) code the extracted data, 3) translate codes into themes, 4) create a model based on the themes, and 5) validate the synthesis. Since the qualitative analysis is exploratory, we do not expect this procedure to be a strict waterfall procedure but steps may be iterated and sometimes merged. An initial coding could for example be done while extracting the relevant data (merging steps one and two) and may need iteration if new codes emerge during the step. For RQ1, our starting point is to identify and code information regarding design decisions when implementing and deploying TRR, while for RQ3 and RQ4 our starting point will be to code effects (indirect and direct) of adopting TRR. 

For RQ1, which is both descriptive and prescriptive, we expect codes and themes to firstly bring insights into which refinements of the general technological rule may be of relevance for a practitioner when selecting an automation strategy and for a researcher interested in investigating the topic further, and secondly to guide positioning of technological rules (prescriptions). Refinements will be expressed in terms of extended taxonomies for the three facets of the technological rule, context, scope, and intervention. These taxonomies will then provide a basis for proposing refined technological rules. In addition, codes and themes that are more context specific (and thus not good candidates for a general theory) will be used to describe the problem instance (our case under study) to support analytical generalization and assessment of the empirical support this case brings to the proposed technological rules. For RQ3 and RQ4, which are more descriptive we expect a more complex model of various effects and their internal relationships. 

Finally, we will assess our interpretations by testing the taxonomies, the technological rules, and the effect model on the study participants.

\section{Quality Assurance and Validity} \label{sec:quality}
The value of design science research may be assessed from three different perspectives~\cite{runeson2020design}, i.e., its \emph{relevance}, its \emph{novelty} and its \emph{rigor}. The design knowledge gained from this research is \emph{relevant} for practitioners facing the challenge of manually assigning bugs to teams, and for researchers studying industrial adoption of ML approaches for automated bug assignment. Relevance is a subjective value~\cite{garousi2020practical} and to support its assessment we will identify and report the context factors that affect the applicability and observed effects of the proposed intervention. Furthermore, the design knowledge is \emph{novel} in terms of increased maturity of the general technological rule and in proposing refined rules with respect to the scope of validity and the effects of adoption. \emph{Rigor} will be achieved by following this preregistered case study protocol and by transparently reporting all steps of interpretation in the qualitative analysis. Rigor may in turn be assessed in terms of construct validity, internal validity, and reliability. As we design a single case study, pure statistical generalisation will not be possible. External validity is instead covered by the discussions on relevance above.

\textbf{Construct Validity.} Since we will conduct an exploratory study, not all constructs will be known upfront. Our high-level constructs such as ``value'' and ``ways of working'' will be refined in the qualitative analysis. The metrics proposed in Figure~\ref{fig:rqs} represent our initial assumptions of how to measure these aspects. We expect the qualitative analysis to reveal additional metrics. To increase the final construct validity, study participants will be asked to assess our interpretations. Finally, we acknowledge that the constructs of TAM have been criticized for being too trivial to result in practical research results. To mitigate this, we will complement TAM's analysis of usefulness  with the quality levels provided by QUPER.

\textbf{Internal Validity.} As discussed in Section~\ref{sec:quan_anal}, we will not be able to perform a controlled randomized trial to prove causal relationships within Ericsson -- we cannot disable HighAuto TRR for a random subset of teams. As we have to deal with the complexity of in vivo research, we aim to conduct a causal Bayesian analysis instead. Still, we will be careful when proposing any causal relationships. To increase the validity of the propositions, confounding factors need to be identified and reported. Some confounders are already known, e.g., product details, organizational structure, and process adaptations, whereas others will emerge from the qualitative analysis.

\textbf{Reliability.} This aspect of rigor concerns to what extent the analysis depends on the specific researchers. We will mitigate threats to reliability through researcher and method \textit{triangulation}~\cite{runeson2012case}. Additional measures include \textit{prolonged involvement}, i.e., the long-term relations evolving during the study will support reliable interpretations, and \textit{member checking}, i.e., participants of the study will validate both data collection and analysis.  


\balance
\bibliographystyle{ACM-Reference-Format}
\bibliography{trr}

\end{document}